\documentclass[preprint,review,3p,12pt]{elsarticle}
\usepackage{amssymb}

\journal{Journal of Crystal Growth}

\begin{document}

\begin{frontmatter}

\title{Crystal growth and characterization of Haldane chain compound
Ni(C$_3$H$_{10}$N$_2$)$_2$NO$_2$ClO$_4$}

\author[1]{W. Tao}
\author[1,2]{L. M. Chen}
\author[1]{X. M. Wang}
\author[1]{C. Fan}
\author[1]{W. P. Ke}
\author[1]{X. G. Liu}
\author[1]{Z. Y. Zhao}
\author[1]{Q. J. Li}
\author[1]{X. F. Sun\corref{cor1}}

\address[1]{Hefei National Laboratory for Physical Sciences at
Microscale, University of Science and Technology of China, Hefei,
Anhui 230026, P. R. China}

\address[2]{Department of Physics, University of Science and
Technology of China, Hefei, Anhui 230026, P. R. China}

\cortext[cor1]{Corresponding author. Tel.: 86-551-3600499, Fax:
86-551-3600499.\\ Email address: xfsun@ustc.edu.cn}


\begin{abstract}

The bulk single crystals of $S = 1$ chain compound
Ni(C$_3$H$_{10}$N$_2$)$_2$NO$_2$ClO$_4$ are grown by using a slow
evaporation method at a constant temperature and a slow cooling
method. It is found that the optimum condition of growing large
crystals is via slow evaporation at 25 $^\circ$C using 0.015 mol
Ni(ClO$_4$)$_2$$\cdot$6H$_2$O, 0.015 mol NaNO$_2$, and 0.03 mol
1,3-propanediamine liquid dissolved into 30 ml aqueous solvent.
High-quality crystals with size up to $18 \times 7.5 \times 5$
mm$^3$ are obtained. The single crystals are characterized by
measurements of x-ray diffraction, magnetic susceptibility,
specific heat and thermal conductivity. The susceptibilities along
three crystallographic axes are found to exhibit broad peaks at
$\sim 55$ K, and then decrease abruptly to zero at lower
temperatures, which is characteristic of a Haldane chain system.
The specific heat and the thermal conductivity along the $c$ axis
can be attributed to the simple phononic contribution and are
analyzed using the Debye approximation.

\end{abstract}


\end{frontmatter}

\newpage

\section{Introduction}

During the past decades, one-dimensional quantum magnets have
attracted an increasing interest because of Haldane's well-known
prediction that for the integer-spin chain system the ground state
is separated from the first excited state by a gap (so-called
Haldane gap), while for the half-integer spin system the
excitation spectrum has no gap \cite{Haldane, Botet, Botet1,
Parkinson, Moreo, Betsuyaku, Nightingale}. The existence of
Haldane gap has a great impact on the low-temperature magnetic
properties and has been verified in lots of one-dimensional
quantum systems, such as $S = 1$ Heisenberg chain coumpounds
CsNiCl$_3$ \cite{Buyers, Steiner} and
Ni(C$_2$H$_8$N$_2$)$_2$NO$_2$(ClO$_4$) \cite{Renard1}, and $S = 2$
quasi-linear chain compound Mn(2, 2'-bipyridine)Cl$_3$
\cite{Perlepes, Granroth}.

Nickel propylene-diamine nitrite perchlorate,
Ni(C$_3$H$_{10}$N$_2$)$_2$NO$_2$ClO$_4$ (NINO), has been found to
be an ideal Haldane chain system for its relatively weak
interchain interaction and obvious magnetic anisotropy
\cite{Renard}. The crystal structure of single crystal NINO was
first reported by Renard \textit{et al.} \cite{Renard}. In NINO,
the $S = 1$ spins of Ni$^{2+}$ are coupled along the $b$ axis with
exchange interaction $J \approx - 50$ K \cite{Takeuchi, Yosida}.
Susceptibility measurement did not show long-range magnetic order
down to 0.5 K, which indicates a negligible interchain exchange
\cite{Renard}. The Haldane gap, uniaxial anisotropy, and
orthorhombic anisotropy parameters were identified as $E_g = 14.2$
K, $D = - 11.5$ K, and $E = 2.1$ K, respectively \cite{Takeuchi}.
Due to the role of crystalline field in NINO, the energy gap is
actually split into three gaps with the values of $\Delta E_1 =
8.3$ K, $\Delta E_2 =12.5$ K, and $\Delta E_3 = 21.9$ K.
\cite{Takeuchi}. When an external magnetic field is applied along
the $a$ axis, $\Delta E_2$ keeps constant, $\Delta E_3$ increases,
and $\Delta E_1$ decreases with increasing field, respectively.
However, due to the existence of staggered transverse field,
$\Delta E_1$ actually descends to a finite value at a critical
field $B_c$, and then increases above $B_c$ \cite{Sieling,
Hagiwara}.

The growth of crystals from aqueous solution is a common method to
synthesize organic single crystals \cite{Meera, Devashankar,
Chen}. The single crystals of NINO were synthesized by several
groups via a slow evaporation method \cite{Renard, Takeuchi,
Sieling, Fukui}. However, the condition of crystal growth and the
sizes of crystals were not depicted in their papers. Therefore, it
is still necessary to make a detailed investigation on the
conditions of crystal growth for NINO. In this work, we try to
grow NINO single crystals by using both a slow evaporation and a
slow cooling method from the aqueous solution. It is found that
the appropriate condition of growing NINO crystals is to dissolve
0.015 molar Ni(ClO$_4$)$_2$$\cdot$6H$_2$O and NaNO$_2$ into 30 ml
aqueous solvent with 0.03 mol 1,3-propanediamine and keep the
solution at 25$^\circ$C for a slow evaporation. The quality of
obtained single crystals is characterized by the x-ray
diffraction, and their physical properties are characterized by
the magnetic susceptibility, specific heat, and thermal
conductivity measurements.

\section{Experimental details}

\subsection{Synthesis and crystal growth}

Ni(C$_3$H$_{10}$N$_2$)$_2$NO$_2$ClO$_4$ is synthesized using
Ni(ClO$_4$)$_2$$\cdot$6H$_2$O, 1,3-propanediamine, and NaNO$_2$ as
raw materials which are mixed with a stoichiometric ratio 1:1:2 in
the aqueous solvent. The chemical reaction equation is

Ni(ClO$_4$)$_2\cdot$6H$_2$O + NaNO$_2$ + 2C$_3$H$_{10}$N$_2$
$\rightarrow$ Ni(C$_3$H$_{10}$N$_2$)$_2$NO$_2$ClO$_4$ + NaClO$_4$
+ 6H$_2$O. 2.5 ml liquid of 1,3-propanediamine (0.03 molar) is
firstly mixed into the deionized water uniformly. Then 0.015 molar
Ni(ClO$_4$)$_2$$\cdot$6H$_2$O and 0.015 molar NaNO$_2$ are added
to the solution slowly with stirring, and the reaction process
occurs immediately, accompanied with releasing heat. The prepared
solution is used to grow single crystals under several different
conditions.

First, we investigate the influence of different degrees of the
initial super-saturation of the solutions on the crystal growth at
a constant temperature 25 $^\circ$C. The pre-prepared tiny
crystals are dropped into the solutions and act as seeds. The
shapes and sizes of the crystals grown using different amounts of
solvent are shown in Fig. 1. It is clear that the effect of the
initial concentration of solution is remarkable on the
crystallization of NINO. The detailed initial conditions at 25
$^\circ$C, growth durations, and sizes of as-grown single crystals
are listed in table 1. It is found that in the 20 ml aqueous
solvent, the mahogany as-grown crystals have a typical dimension
of (3.5--6.5) $\times$ (3.5--4.5) $\times$ (1.5--2.5) mm$^3$. From
30 ml aqueous solvent, we obtain very large single crystals with
size up to (16--18) $\times$ (7.5--8) $\times$ (3.5--5) mm$^3$ and
excellent crystallographic surfaces, as shown in Fig. 1(b).
However, when the volume of solvent is larger than 30 ml, the
sizes of as-grown crystals become smaller with gradually
increasing the amount of solvent. It is known that the crystal
growth from an aqueous solution is based on the existence of
metastable supersaturation regions where spontaneous nucleation is
impossible, so it is possible to grow a big single crystal from a
seed \cite{Laudise}. For the current case, the degree of the
initial super-saturation decreases with increasing the amounts of
solvent under the condition that the amount of raw material is
invariant at the given temperature 25 $^\circ$C. It is likely that
the concentration of the solution with 30 ml aqueous solution is
just under or close to the metastable state making the
crystallization occur continuously on seed crystals. Although
increasing the super-saturation can accelerate the crystal growth,
at a high super-saturation degree there is a strong tendency for
crystals to nucleate spontaneously, {\it i.e.}, to produce many
small crystals at the same time rather than to form a single large
one.

Based on the above results, we further investigate the influence
of the temperature on the crystal growth in 30 ml aqueous solvent
with the same amount of raw materials as the above. The
morphological features of the obtained crystals are shown in Fig.
2. The detailed temperature conditions, growth periods, and sizes
of as-grown single crystals are listed in table 2. It turns out
that when the temperature of the solution is above 60 $^\circ$C,
only a large amount of polycrystals are obtained and no sizeable
single crystals can be formed. When the solution is kept at 35
$^\circ$C, large single crystals with the typical size of
(13.5--15) $\times$ (4--4.5) $\times$ (3.0--4.0) mm$^3$ and bright
surfaces are obtained. The sizes of single crystals become smaller
with increasing temperature, as shown in Fig. 2. The reason can be
as following: the solubility of NINO in the aqueous solvent
increases as the temperature increases, which thus decreases the
degree of super-saturation; on the other hand, increasing
temperature enforces the driving force of the liquid-to-solid
phase transition and thus accelerates the process of the crystal
growth by enhancing the diffusions of the ions to the surface.
However, a higher temperature simultaneously increases the energy
diffusion of ions, and too high temperature makes diffusion too
fast to form single crystals, just as the case of 60 $^\circ$C
growth indicates \cite{Chen, Singh}.

In addition, a slow cooling method is also carried out to grow
NINO single crystals. The slow cooling process of solvent with 30
ml deionized water is performed from 65 to 25 $^\circ$C at a rate
of 5 $^\circ$C per day. The results are shown in Fig. 3. It can be
seen that although single crystals as large as 11 $\times$ 5
$\times$ 3.5 mm$^3$ can be formed, they have no regular shapes and
clean cleavage planes, indicating that the present condition is
not the most appropriate for the crystal growth of NINO. Actually,
the slow cooling method is a much more proper way to grow those
single crystals that have large temperature coefficients of
solubilities, which also needs well-controlled temperature
descending-rate.

To sum up, we compare the results of single crystal growth of NINO
under different conditions, including both a slow evaporation
method and a slow cooling method. It turns out that the slow
evaporation at 25 $^\circ$C in the 30 ml aqueous solvent with the
solutes of 0.015 mol Ni(ClO$_4$)$_2$$\cdot$6H$_2$O and 0.015 mol
NaNO$_2$ as well as 0.03 mol 1,3-propanediamine is the most
appropriate condition to grow the high-quality NINO single
crystals. In passing, it is notable that compared with the growth
of (CH$_3$)$_2$NH$_2$CuCl$_3$ (MCCL for short) single crystals
using the slow evaporation method \cite{Chen}, the growth of NINO
crystals shows some similar features with MCCL; for example, the
sizes of crystals are greatly affected by the initial
concentrations of the solutions, and it is difficult to get single
crystals in the solutions at too high temperature.

\subsection{Characterization}

Powder x-ray diffraction measurements are carried out using an
X'Pert PRO X-Ray diffractometer (PANalytical Company).
Magnetization measurement is performed in the temperature range of
2 -- 300 K in an applied field of 1 T using a commercial SQUID-VSM
(Quantum Design) and specific heat is measured using a PPMS
equipped with a $^3$He refrigerator (Quantum Design). The thermal
conductivity with heat flow along the $c$ axis ($\kappa_c$) is
measured using a conventional steady-state technique. The data of
$\kappa_c$ at 4 -- 100 K are obtained using a pulsed-tube
refrigerator and the low-$T$ data at 0.3 -- 8 K are taken using a
$^3$He refrigerator, with the temperature gradient detected by
using a Chromel-Constantan thermocouple and two RuO$_2$ chip
resistors, respectively \cite{Sun, Wang}.

\section{Results and discussion}

\subsection{X-ray diffraction}

Figure 4 shows the result of x-ray powder diffraction at room
temperature, in which the indices of diffraction peaks are
determined according to the orthorhombic structure (space group
$Pbn2_1$) with lattice parameters $a$ = 15.384 {\AA}, $b$ = 10.590
{\AA} and $c$ = 8.507 {\AA} \cite{Renard}. We also use the x-ray
diffraction to determine the crystallographic axes and the results
clearly indicate that the as-grown crystals are grown along the
$c$ axis and the largest cleavage surface of crystals is the (100)
plane, which coincide with the earlier results \cite{Takeuchi}.
The rocking curve of the (200) reflection is shown in Fig. 5(a),
and the full width at half maximum (FWHM) is only 0.075$^\circ$,
indicative of the high quality of the single crystal. Figure 5(b)
shows the ($h$00) diffraction pattern of this crystal.

\subsection{Magnetic susceptibility and magnetization}

Figure 6(a) shows the temperature dependences of magnetic
susceptibilities in 1 T field applied along three crystallographic
axes $a$, $b$, and $c$, respectively, which are essentially
consistent with the previous measurement \cite{Takeuchi}. It is
clearly seen that three magnetic susceptibilities exhibit rounded
peaks at $\sim$ 55 K, and fall down abruptly to zero at lower
temperatures, which reflects the excitation gap of NINO and is
characteristic of the Haldane chain. Note that the effect of the
crystalline field is remarkable only at very low temperatures. It
is applicable to employ the theoretical calculation for an
isotropic Heisenberg chain by the exact diagonalization at
relatively high temperature, which is given by \cite{Meyer}
\begin{equation}
\chi=\frac{Ng^2\mu_B^2}{k_BT}
\left[\frac{2+0.01947x+0.777x^2}{3+4.346x+3.232x^2+5.834x^3}\right],
\label{chi}
\end{equation}
where $\chi$ is the molar susceptibility, $x=|J|/k_BT$, $N$ the
number of Ni$^{2+}$ ions per mole NINO, $g$ the Land\'e factor,
$\mu_B$ the Bohr magneton, and $k_B$ the Boltzmann constant. As
shown in Fig. 6(a), the data at high temperature between 50 and
300 K are well fitted using the above equation, with the
parameters $J = - 50.0$ K, $g_a = 2.42$, $g_b = 2.25$, and $g_c =
2.37$. The fitting parameters are found to be in good consistent
with those in some former works \cite{Meyer}.

The magnetization curves of NINO along three principal axes $a$,
$b$, and $c$ at $T$ = 2 K are shown in Fig. 6(b). It is observed
that the magnetization process shows an obvious anisotropic
behavior along three different crystallographic axes. The
magnetization along the $a$ axis shows a slightly faster slope
than that along the $c$ axis, and in contrast the magnetization of
the $b$ axis exhibits a rather steep increase with increasing
magnetic field. In spite of these obvious distinctions, a
noticeable feature of the $M(H)$ curves is that the magnetization
is in fact rather small. The reason is obvious, that is, at low
temperatures the system is nonmagnetic because of the presence of
the Haldane gap \cite{Katsumata}. Only when an applied field
quenches the energy gap, the system becomes magnetic. Actually,
for NINO the magnetic field needed to close the gap is rather
high, about 9 T \cite{Takeuchi}.

\subsection{Specific heat and thermal conductivity}

Figure 7(a) shows the specific heat at low temperatures, which do
not show any sign of phase transition. To quantitatively analyze
the data, we use the simple Debye approximation of the phonon
specific heat \cite{Tari}
\begin{equation}
C_p = \beta T^3 + \gamma T^5 + \delta T^7, \label{C}
\end{equation}
where $\beta =(12{\pi}^4nk_B/5{\Theta}_D^3)$ with $n$ the number
density of the atoms in NINO and $\Theta_D$ the Debye temperature;
$\gamma$ and $\delta$ are temperature-independent coefficients.
Note that the high-power terms are included because at not very
low temperatures the density of phonon modes of a solid actually
deviates from the $\omega^2$-law of Debye's presupposition. Figure
7(a) shows that the experimental data can be well fitted to Eq.
(\ref{C}), indicating that the contribution of the specific heat
is completely due to the lattice vibration. The corresponding
fitting parameters are $\beta$ = 5.74 $\times$ 10$^{-3}$ J/K$^4$
mol, $\gamma = - 1.24$ $\times$ 10$^{-5}$ J/K$^6$ mol, and
$\delta$ = 1.17 $\times$ 10$^ {-8}$ J/K$^8$ mol. The Debye
temperature $\Theta_D$ = 236.4 K is calculated using the fitting
parameter $\beta$.

Figure 7(b) shows the temperature dependence of the thermal
conductivity along the $c$ axis of a crystal with size of 3.7
$\times$ 0.6 $\times$ 0.4 mm$^3$. It is seen that a pronounced
phononic peak is observed at about 6 K. Here we analyze $\kappa_c$
using the Debye approximation for the phonon spectrum and
relaxation approximation, where the phonon thermal conductivity is
expressed as \cite{Berman}
\begin{equation}
\kappa_{ph}=\frac{k_B}{2\pi^2v}(\frac{k_B} {\hbar})^3T^3
\mbox{\resizebox{1.5\width}{1.5\height}{$\int$}}_0^{\Theta_D / T}
\frac{x^4e^x}{(e^x-1)^2}\tau(\omega,T)dx, \label{kappa}
\end{equation}
in which $x = \hbar \omega/k_BT$ is dimensionless, $\omega$ is the
phonon frequency, and $1/\tau(\omega,T)$ is the phonon-scattering
relaxation rate. The average sound velocity $v$ = 1644 m/s is
calculated from the Debye temperature using the formula $\Theta_D
= v(\hbar/k_B)(6\pi^2n)^{1/3}$. The relaxation time takes three
terms as follows
\begin{equation}
\tau^{-1} = \upsilon/L + A\omega^4 + BT\omega^3\exp(-\Theta_D/bT),
\label{tau}
\end{equation}
which represent the phonon scattering by the grain-boundary,
point-defect, and phonon-phonon Umklapp scattering, respectively,
with $L$, $A$, $B$ and $b$ free parameters. The thermal
conductivity can be well fitted to Eq. (\ref{tau}) up to 70 K, as
shown in Fig. 7(b). The fitting parameters are $L$ =
7.1$\times$10$^{-5}$ m, $A$ = 2.0$\times$10$^{-41}$ s$^3$, $B$ =
9.6$\times$10$^{-29}$ K$^{-1}$s$^2$, and $b$ = 10.2. These
indicates that although NINO is a magnetic insulator, the magnetic
excitations make no contribution to $\kappa_c$ because the heat
current is perpendicular to the spin-chain direction.

\section{CONCLUSIONS}

Different growth conditions for a slow evaporation method and a
slow cooling method are used to grow NINO single crystals. It is
found that the optimum condition of growing large single crystals
is via the slow evaporation at constant temperature of 25
$^\circ$C using the raw materials (0.015 mol
Ni(ClO$_4$)$_2$$\cdot$6H$_2$O, 0.015 mol NaNO$_2$, and 0.03 mol
1,3-propanediamine liquid) dissolved into 30 ml aqueous solvent.
X-ray diffraction data have confirmed the good crystallinity of
the grown crystals. The magnetic susceptibilities are found to
show broad maximum at about 55 K, which is characteristics of the
1D Haldane chain system, and the higher-$T$ data of $\chi$ are
analyzed with the isotropic Heisenberg model. The low-temperature
specific heat and the $c$-axis thermal conductivity are found to
be purely phononic behaviors and can be well described by the
Debye model.

\section*{ACKNOWLEDGMENTS}

This work was supported by the Chinese Academy of Sciences, the
National Natural Science Foundation of China, and the National
Basic Research Program of China (Grant Nos. 2009CB929502 and
2011CBA00111).

\bibliographystyle{elsarticle-num}
\bibliography{<your-bib-database>}

\begin{thebibliography}{}

\bibitem{Haldane}
F. D. M. Haldane, Phys. Lett. A {\bf 93}, 464 (1983); Phys. Rev.
Lett. {\bf 50}, 1153 (1983).

\bibitem{Botet}
R. Botet and R. Jullien, Phys. Rev. B {\bf 27}, 613 (1983).

\bibitem{Botet1}
R. Botet, R. Jullien, and M. Kolb, Phys. Rev. B {\bf 29}, 5216
(1983).

\bibitem{Parkinson}
J. B. Parkinson and J. C. Bonner, Phys. Rev. B {\bf 32}, 4703
(1985).

\bibitem{Moreo}
A. Moreo, Phys. Rev. B {\bf 35}, 8562 (1987).

\bibitem{Betsuyaku}
H. Betsuyaku, Phys. Rev. B {\bf 36}, 799 (1987).

\bibitem{Nightingale}
M. P. Nightingale and H. W. Bl\"{o}te, Phys. Rev. B {\bf 33}, 659
(1986).

\bibitem{Buyers}
W. J. L. Buyers, R. M. Morra, R. L. Armstrong, M. J. Hogan, P.
Gerlach, and K. Hirakawa, Phys. Rev. Lett. {\bf 56}, 371 (1986).

\bibitem{Steiner}
M. Steiner, K. Kakurai, J. K. Kjems, D. Petitgrand, and R. Pynn,
J. Appl. Phys. {\bf 61}, 3953 (1987).

\bibitem{Renard1}
J. P. Renard, M. Verdaguer, L. P. Regnault, W. A. C. Erkelens, J.
Rossat-Mignod, and W. G. Stirling, Europhys. Lett. {\bf 3}, 945
(1987).

\bibitem{Perlepes}
S. P. Perlepes, A. G. Blackman, J. C. Huffman, and G. Christou,
Inorg. Chem. {\bf 30}, 1665 (1991).

\bibitem{Granroth}
G. E. Granroth, M. W. Meisel, M. Chaparala, Th. Jolicoeur, B. H.
Ward, D. R. Talham, Phys. Rev. Lett. {\bf 77}, 1616 (1996).

\bibitem{Renard}
J. P. Renard, M. Verdaguer, L. P. Regnault, W. A. C. Erkelens, J.
Rossat-Mignod, J. Ribas, W. G. Stirling, and C. Vettier, J. Appl.
Phys. {\bf 63}, 3538 (1988).

\bibitem{Takeuchi}
T. Takeuchi, M. Ono, H. Hori, T. Yosida, A. Yamagishi, and M.
Date, J. Phys. Soc. Jpn. {\bf 61}, 3255 (1992).

\bibitem{Yosida}
T. Yosida and M. Fukui, J. Phys. Soc. Jpn. {\bf 61}, 2304 (1992).

\bibitem{Sieling}
M. Sieling, U. L\"{o}w, B. Wolf, S. Schmidt, S. Zvyagin, and B.
L\"{u}thi. Phys. Rev. B {\bf 61}, 88 (2000).

\bibitem{Hagiwara}
M. Hagiwara and K. Katsumata, Phys. Rev. B {\bf 53}, 14319 (2000).

\bibitem{Meera}
K. Meera, R. Muralidharan, R. Dhanasekaran, Prapun Manyum, and P.
Ramasamy, J. Cryst. Growth {\bf 263}, 510 (2004).

\bibitem{Devashankar}
S. Devashankar, L. Mariappan, P. Sureshkumar, and M. Rathnakumari,
J. Cryst. Growth {\bf 311}, 4207 (2009).

\bibitem{Chen}
L. M. Chen, W. Tao, Z. Y. Zhao, Q. J. Li, W. P. Ke, X. G. Liu, C.
Fan, and X. F. Sun, J. Cryst. Growth {\bf 312}, 3243 (2010).

\bibitem{Fukui}
M. Fukui and T. Yosida, J. Phys. Soc. Jpn. {\bf 60}, 739 (1991).

\bibitem{Laudise}
R. A. Laudise, {\it The Growth of Single Crystal} (Prentice-Hall,
Inc., Englewood Cliffs, New Jersey, 1970).

\bibitem{Singh}
N. B. Singh, T. Henningsen, R. H. Hopkins, R. Mazelsky, R. D.
Hamacher, E. P. Supertzi, F. K. Hopkins, D. E. Zelmon, and O. P.
Singh, J. Cryst. Growth {\bf 128}, 976 (1993).

\bibitem{Sun}
X. F. Sun, W. Tao, X. M. Wang, and C. Fan, Phys. Rev. Lett. {\bf
102}, 167202 (2009).

\bibitem{Wang}
X. M. Wang, C. Fan, Z. Y. Zhao, W. Tao, X. G. Liu, W. P. Ke, X.
Zhao, and X. F. Sun. Phys. Rev. B {\bf 82}, 094405 (2010).

\bibitem{Meyer}
A. Meyer, A. Gleizes, J. Girerd, M. Verdaguer, and O. Kahn, Inorg.
Chem. {\bf 21} 1729 (1982).

\bibitem{Katsumata}
K. Katsumata, H. Hori, T. Takeuchi, M. Date, A. Yamagishi, and J.
P. Renard, Phys. Rev. Lett. {\bf 63}, 86 (1989).

\bibitem{Tari}
A. Tari, {\it Specific Heat of Matter at Low Temperatures}
(Imperial College Press, 2003).

\bibitem{Berman}
R. Berman, {\it Thermal Conduction in solids} (Oxford University
Press, Oxford, 1976).

\end{thebibliography}

\newpage

\begin{figure}
\includegraphics[clip,width=8cm]{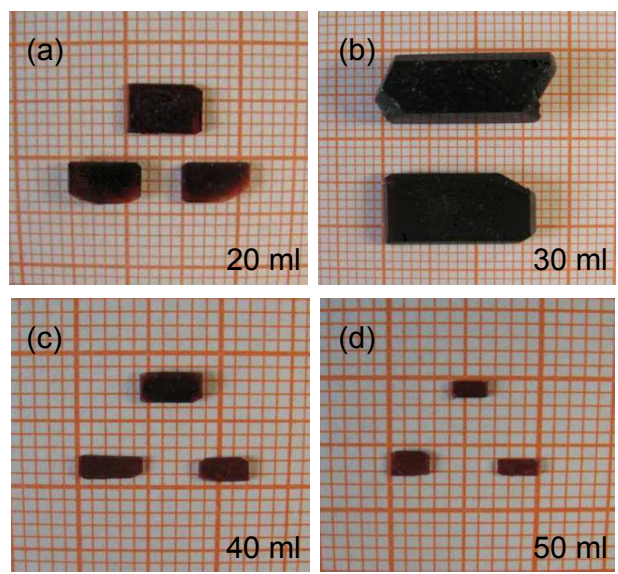}
\caption{NINO single crystals grown from different amounts of
aqueous solvents at 25 $^\circ$C. Photos (a), (b), (c), and (d)
correspond to the 20, 30, 40, and 50 ml aqueous solvent,
respectively.}
\end{figure}

$ $
\newpage

\begin{figure}
\includegraphics[clip,width=12cm]{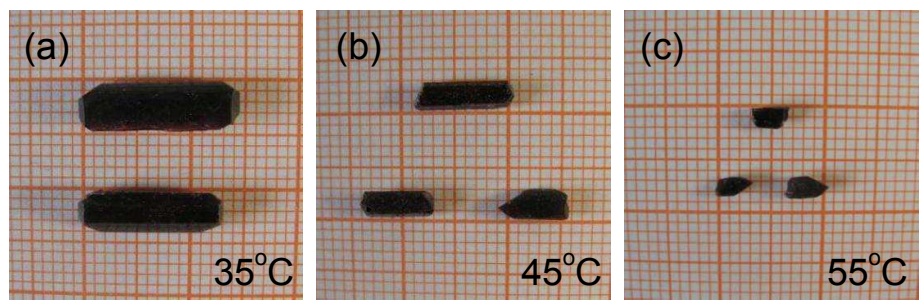}
\caption{NINO single crystals grown from the 30 ml solvents at the
temperatures of 35, 45, and 55 $^\circ$C, respectively.}
\end{figure}

$ $
\newpage

\begin{figure}
\includegraphics[clip,width=5cm]{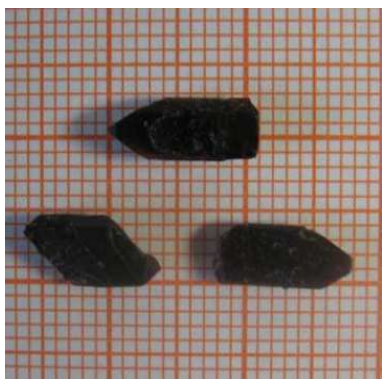}
\caption{NINO single crystals grown from 30 ml solvent with
temperature decreasing slowly from 65 to 25 $^\circ$C at a rate of
5 $^\circ$C per day.}
\end{figure}

$ $
\newpage

\begin{figure}
\includegraphics[clip,width=12cm]{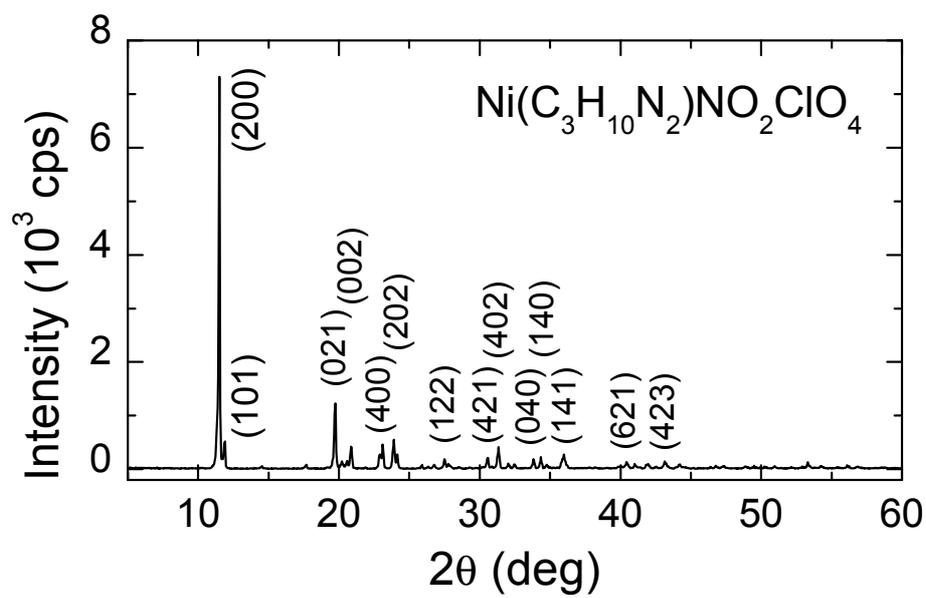}
\caption{ X-ray powder diffraction pattern of NINO at room
temperature.}
\end{figure}

$ $
\newpage

\begin{figure}
\includegraphics[clip,width=12cm]{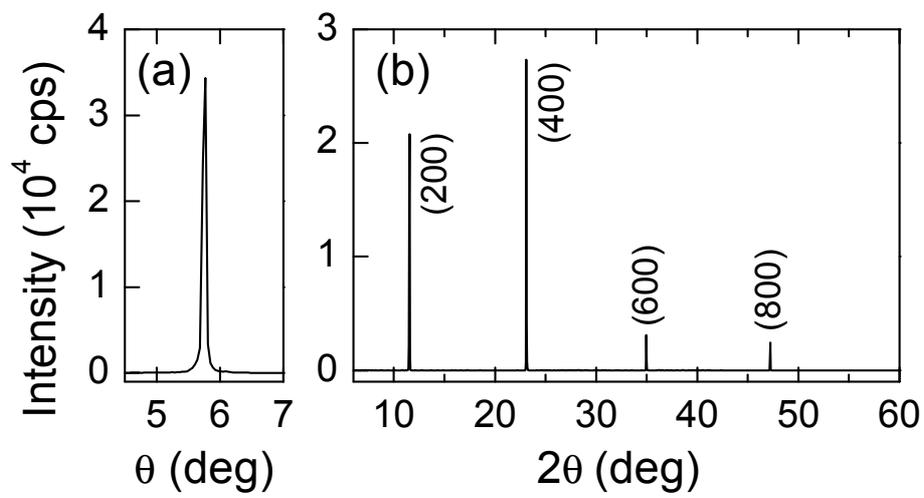}
\caption{(a) X-ray rocking curve of the (200) reflection, with a
FWHM of only 0.075$^\circ$. (b) ($h$00) diffraction pattern.}
\end{figure}

$ $
\newpage

\begin{figure}
\includegraphics[clip,width=8cm]{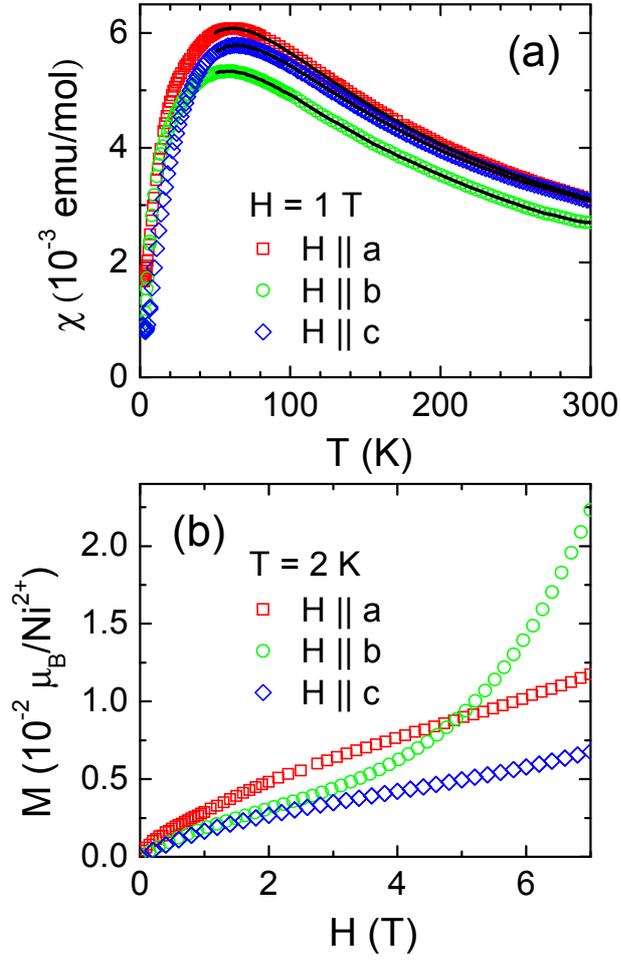}
\caption{(a) Temperature dependences of magnetic susceptibilities
in 1 T field applied along three crystallographic axes $a$, $b$,
and $c$, respectively. The solid lines show the theoretical
fitting curves in the temperature range between 50 and 300 K (see
text). (b) Magnetic-field dependence of magnetization at $T = 2$
K.}
\end{figure}

$ $
\newpage

\begin{figure}
\includegraphics[clip,width=8cm]{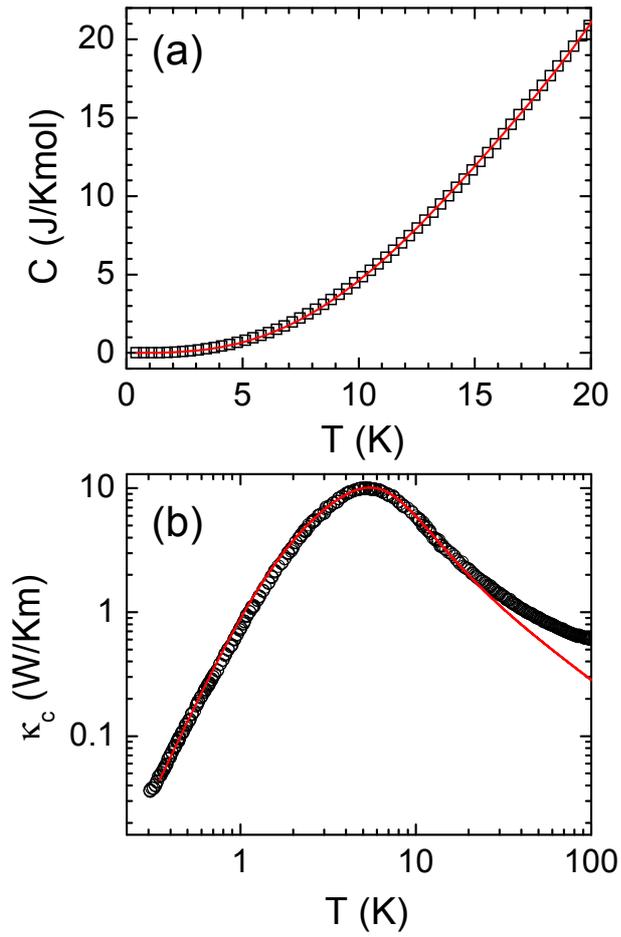}
\caption{(a) Low-temperature specific heat of NINO single crystal.
The solid line represents the theoretical fitting curve (see
text). (b) Temperature dependence of thermal conductivity with
heat current along the $c$ axis. The solid line represents the
fitting curve based on the Debye model for the phononic heat
transport (see text).}
\end{figure}

\clearpage

\newpage


\begin{table}[htbp]
\caption {Results for different volumes of aqueous solution kept
at 25 $^\circ$C. The raw materials are 0.015 mol
Ni(ClO$_4$)$_2$$\cdot$6H$_2$O and 0.015 mol NaNO$_2$ as well as
0.03 mol 1,3-propanediamine.} \centering
\begin{tabular}{p{80pt}p{80pt}p{280pt}}
\hline \hline
Volume (ml) &  Growth Period (days) & Crystal State  \\
\hline
20  & 34   & big crystals, size about (3.5--6.5)$\times$(3.5--4.5)$\times$(1.5--2.5) mm$^3$ \\
30  & 58   & large bright crystals, size about (16--18)$\times$(7.5--8)$\times$(3.5--5) mm$^3$\\
40  & 62   & small crystals, size about (3.5--4.5)$\times$(1--2)$\times$(0.2--0.4) mm$^3$ \\
50  & 65   & small thin pieces, size about (2.0--3)$\times$(1.2--2.2) mm$^2$ \\
\hline \hline
\end{tabular}
\end{table}

\clearpage

\newpage


\begin{table}[htbp]
\caption {Results for different growing temperatures in 30 ml
aqueous solution. The raw materials are 0.015 mol
Ni(ClO$_4$)$_2$$\cdot$6H$_2$O and 0.015 mol NaNO$_2$ as well as
0.03 mol 1,3-propanediamine.} \centering
\begin{tabular}{p{80pt}p{80pt}p{280pt}}
\hline \hline
Temperature ($^\circ$C) & Growth Period (days)& Crystal State  \\
\hline
35  & 32   & large bright crystals, size about (13.5--15)$\times$(4--4.5)$\times$(3.0--4.0) mm$^3$\\
45  & 20   & big bright crystals, size about (6.5--9)$\times$(2--3)$\times$2.5 mm$^3$\\
55  & 7    & small irregular crystals \\
60  & 5    & none \\
\hline \hline
\end{tabular}
\end{table}

\clearpage

\end{document}